\documentclass[12pt]{iopart}   
\usepackage{graphicx}
\usepackage{epsfig}
\usepackage{amssymb}
\usepackage{array}

\newcommand{\be}{\begin{equation}}
\newcommand{\ee}{\end{equation}}
\newcommand{\beqn}{\begin{eqnarray}}
\newcommand{\eeqn}{\end{eqnarray}}

\begin{document}

\title{Critical dynamics of the Kuramoto model on sparse random networks}
\author{R\'obert Juh\'asz}
\ead{juhasz.robert@wigner.mta.hu}
\address{Wigner Research Centre for Physics, Institute for Solid State
Physics and Optics, H-1525 Budapest, P.O.Box 49, Hungary}
\author{Jeffrey Kelling}
\ead{j.kelling@hzdr.de}
\address{Department of Information Services and Computing,
Helmholtz-Zentrum Dresden-Rossendorf,
P.O.Box 51 01 19, 01314 Dresden, Germany}
\author{G\'eza \'Odor}
\ead{odor@mfa.kfki.hu}
\address{Institute of Technical Physics and Materials Science,
Centre for Energy Research of the Hungarian Academy of Sciences,
P.O.Box 49, H-1525 Budapest, Hungary}
\date{\today}

\begin{abstract}
We consider the Kuramoto model on sparse random networks such as the Erd\H os-R\'enyi graph or its combination with a regular two-dimensional lattice and 
study the dynamical scaling behavior of the model at the synchronization transition by large-scale, massively parallel numerical integration. 
By this method, we obtain an estimate of critical coupling strength  more accurate than obtained earlier by finite-size scaling of the stationary order parameter. Our results confirm the compatibility of the correlation-size and the temporal correlation-length exponent with the mean-field universality class.  
However, the scaling of the order parameter exhibits corrections much stronger than those of the Kuramoto model with all-to-all coupling, making thereby an accurate estimate of the order-parameter exponent hard.   
We find furthermore that, as a qualitative difference to the model with all-to-all coupling, the effective critical exponents involving the order-parameter exponent, such as the effective decay exponent characterizing the critical desynchronization dynamics show a non-monotonic approach toward the asymptotic value. 
In the light of these results, the technique of finite-size scaling of limited size data for the Kuramoto model on sparse graphs has to be treated cautiously. 
\end{abstract}

\maketitle

\section{Introduction}

Synchronization of parts of interacting systems is a ubiquitous phenomenon in nature. It is observed in biological, chemical, physical, and sociological systems and much effort has been devoted to the theoretical understanding of its general features \cite{pikovsky,acerbon,arenas}.  
A paradigmatic model of $N$ globally coupled oscillators was introduced and solved in the stationary state in the limit $N\to\infty$ by Kuramoto \cite{kuramoto}, and later the macroscopic evolution of the system was shown to be governed by a finite set of nonlinear ordinary differential equations \cite{oa}. 
An interesting property of the Kuramoto model is that it shows a continuous phase transition with a diverging correlation size, separating a synchronized phase from an unsynchronized one. Owing to its chaotic time evolution, it obeys a scaling theory analogous to critical points of stochastic systems. Here, the whole set of critical exponents are known \cite{kuramoto,oa,chate_prl}, and the corresponding universality class is termed as mean field since, due to the all-to-all coupling, the individual oscillators interact with a mean field of the rest of the oscillators.  

A challenging line of research aims at studying the possibility and nature of phase synchronization transition in extended systems, where oscillators sit on a regular lattice of finite dimension $d$ and the interaction, in the extreme case, is restricted to nearest-neighbors \cite{chate_prl,sakaguchi,hong2005}, or, as an interpolation, its strength decays with the distance between oscillators \cite{rogers,vicsek,chowdhury,uchida}. Another intermediate case between finite-dimensional systems and all-to-all coupling are sparse random networks whose diameter scales with the size slower than a power law \cite{hong2002,um}.
In these systems, exact tractability is, in general, lost and one resorts to approximations or numerical integration. 
Using linearization of the dynamical equations of the Kuramoto oscillators, which is justified by the system being in the synchronized phase, consistency, i.e. the existence of synchronization, has been found in regular lattices for dimensions $d>4$ \cite{hong2005}, while, for general networks, for $d_s>4$, where $d_s$ is the spectral dimension of the network \cite{bianconi}. 
Another question is the nature of the synchronization transition, provided it exists, for which numerical results are available, mainly on the scaling of the order parameter in the stationary state. 
For regular lattices of dimensions $d=5,6$, which are above the lower critical dimension $d_l=4$ \cite{chate_prl}, as well as for random graphs with an infinite spectral dimension such as Erd\H os-R\'enyi (ER) graphs \cite{um} or Watts-Strogatz networks \cite{lee}, compatibility with the mean-field universality class was found. 
Besides the pure theoretical interest, studying this problem is greatly motivated by neuroscience, since brain is expected to work near a synchronization transition \cite{MunPNAS}.

In this work, we revisit the Kuramoto model on Erd\H os-R\'enyi graphs, but instead of restricting ourselves to the static scaling, we also study numerically the dynamical scaling, first time in this model, as it has been done for the Kuramoto model with all-to-all coupling in Ref. \cite{choi}. 
Furthermore, we also consider the Kuramoto model on random networks which are combinations of ER graphs and a regular, two-dimensional (2d) lattice. 
A motivation for studying such structures is that certain applications of the model contain random, long-range connections embedded in Euclidean space where they compete with the effect of short-range interactions. Such a situation is typical, for example, in brain networks \cite{HMN}. 

The numerical integration of the dynamical equtions has been performed on graphics processing units, allowing us to reach size and time scales much larger than those applied so far. 
In earlier works, the mean-field behavior of systems with $d_s>4$  was usually demonstrated by finite-size scaling of the stationary order parameter using the mean-field critical exponents and tuning the critical coupling strength to achieve an optimal collapse. But this method, at least for the limited system sizes that are numerically available, is rather uncertain and gives a satisfactory scaling collapse for a whole range of combinations of parameters. 
Therefore we used instead the dynamical scaling of the time-dependent order parameter to estimate the critical coupling strength. 
According to our numerical results, the correlation size and the temporal correlation-length exponent are compatible with the mean-field universality class, however, the sparse interaction network leads to 
strong corrections to the scaling of the time-dependent order parameter. The effective, time-dependent decay exponent $\delta_{\rm eff}(t)$ characterizing the critical desynchronization dynamics shows a qualitatively different dependence on time compared to the model with all-to-all coupling: 
Rather than a monotonically increasing behavior, it exhibits an initial increase followed by a decreasing approach to the asymptotical value. 
Accordingly, the asymptotical scaling region is shifted to time and size scales much larger than those of the model with all-to-all coupling, thereby making an accurate estimate of the decay exponent hard. Using size and time scales in the strong-correction-region the estimates of critical exponents involving the order-parameter exponent can significantly differ from the asymptotical values. Conversely, setting the critical exponents to their mean-field values and estimating the critical coupling strength by finite-size scaling can give a considerable error. 
   
The rest of the paper is organized as follows. In sec. \ref{sec:model}, the model is defined and the scaling theory of the synchronization transition is recapitulated. Numerical results for ER graphs and their union with a regular 2d lattice are presented in sec. \ref{sec:numerical} and discussed in sec. \ref{sec:discussion}. 

%%%%%%%%%%%%%%%%%%%%%%%%%%%%%%%%%%%%%%%%%%%%%%%%%%%%%%%%%%%%%%%%%%%%%%%%%%%%
\section{The model}
\label{sec:model} 

We consider the Kuramoto model of interacting oscillators sitting at the nodes of a network, whose phases $\phi_j(t)$, $j=1,2\dots,N$, evolve according to the following set of dynamical equations
\be 
\frac{d\phi_j(t)}{dt}=\omega_j+K\sum_k\sin[\phi_k(t)-\phi_j(t)].
\label{diffeq}
\ee
Here, $\omega_j$  is the intrinsic frequency of the $j$th oscillator, which is drawn from a Gaussian distribution with zero mean and unit variance, and the summation is performed over adjacent nodes of node $j$. 
We are interested in the properties of the phase synchronization transition through studying the phase order parameter defined by
\be
R(t)=\frac{1}{N}\left|\sum_{j=1}^Ne^{i\phi_j(t)}\right|,
\label{op}
\ee
which is non-zero above a critical coupling strength, $K>K_c$, and zero for $K<K_c$ in the limit $N\to\infty$ and $t\to\infty$.
In a finite system of size $N$ and close to the critical point, i.e. for 
$|\Delta|\ll 1$, where $\Delta\equiv K-K_c$, the order parameter averaged over different realizations of the intrinsic frequencies and random networks is expected to have the following scaling property
\be 
R(t,\Delta,N)=b^{-\beta/\overline{\nu}}\tilde R(tb^{-\overline{z}},\Delta b^{-1/\overline{\nu}},Nb^{-1}),
\label{scaling}
\ee
where $b$ is a scale factor, $\beta$ is the order parameter exponent, and the dynamical exponent $\overline{z}$ is related to the correlation-size exponent $\overline{\nu}$ and temporal correlation length exponent $\nu_{\parallel}$ through
$\overline{z}=\nu_{\parallel}/\overline{\nu}$. Here, the scaling function $\tilde R$ is different for different initial states of the system. 
If the system starts from a fully synchronized state [$R(t=0)=1$], then at the critical point ($\Delta =0$) we obtain by setting $b=t^{1/\overline{z}}$ in Eq. (\ref{scaling}) an algebraic decay 
\be 
R(t)\sim t^{-\delta}, 
\label{delta}
\ee
with $\delta=\beta/\nu_{\parallel}$, for times $1\ll t\ll N^{\overline{z}}$.  
In the case of a random initial state, in which $R(t=0)=O(N^{-1/2}$), we have 
an algebraic increase of the order parameter in the critical point according to
\be 
R(t,N)\sim N^{-1/2}t^{\Theta},
\ee
where $\Theta=(1/2-\beta/\overline{\nu})/\overline{z}$, for times $1\ll t\ll N^{\overline{z}}$ \cite{choi}.

The mean-field universality class is characterized by the critical exponents 
$\beta=1/2$ \cite{kuramoto}, $\nu_{\parallel}=1$ \cite{oa,choi}, and $\overline{\nu}=5/2$ \cite{chate_prl}, so the dynamical critical exponents take the values 
$\overline{z}=2/5$, $\delta=1/2$, and $\Theta=3/4$.

%%%%%%%%%%%%%%%%%%%%%%%%%%%%%%%%%%%%%%%%%%%%%%%%%%%%%%%%%%%%%%%%%%%%%%%%%%%
\section{Numerical results}
\label{sec:numerical} 

We numerically integrated the dynamical equations (\ref{diffeq}) on different networks, starting with either equal, or uniformly distributed random phases. We calculated the order parameter given in Eq. (\ref{op}) as a function of time and performed an average over typically a few hundreds of realizations of random networks and intrinsic frequencies. For the numerical integration, we applied the fourth-order Runge-Kutta method with a step length $\Delta t=0.1$, and convinced ourselves that a further reduction of the step length would not noticeably alter the average order parameter. In order to achieve the large system sizes required to study the time-dependence
of the order parameter in the transient regime, we performed part of the
simulations on graphics processing units~(GPUs), increasing simulation
throughput by up to a factor of $80$ for ER graphs with $N=2^{27}$ nodes. 
The details of our GPU implementation will be published elsewhere \cite{KOGkuramotoGPU_tbp}. 

For the studying of the time-dependence at the critical point we found it useful to calculate effective time-dependent exponents from $R(t)$ such as $\delta_{\rm eff}(t)$ in the case of a synchronized initial state: 
\be 
\delta_{\rm eff}(t)=-\frac{\ln[R(t')/R(t)]}{\ln(t'/t)}, 
\ee
where we used typically $\ln(t'/t)=0.23$. 

First, we considered random graphs with a constant number of links, $E=\frac{k}{2}N$, in each sample of size $N$, which connect equiprobably chosen pairs of nodes. The numerical calculations were performed on random graphs with a mean degree $k=4$.  

\begin{figure}
\begin{center}
\includegraphics[width=10cm]{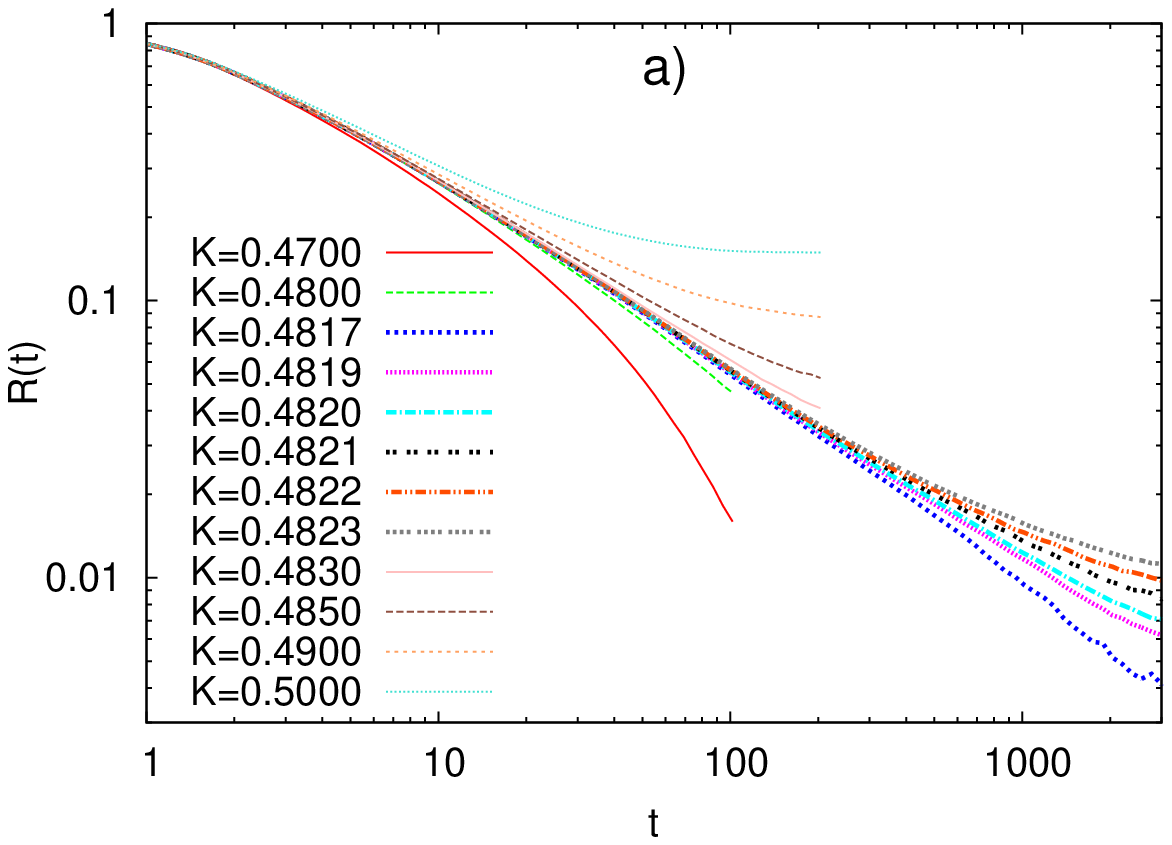} 
\includegraphics[width=10cm]{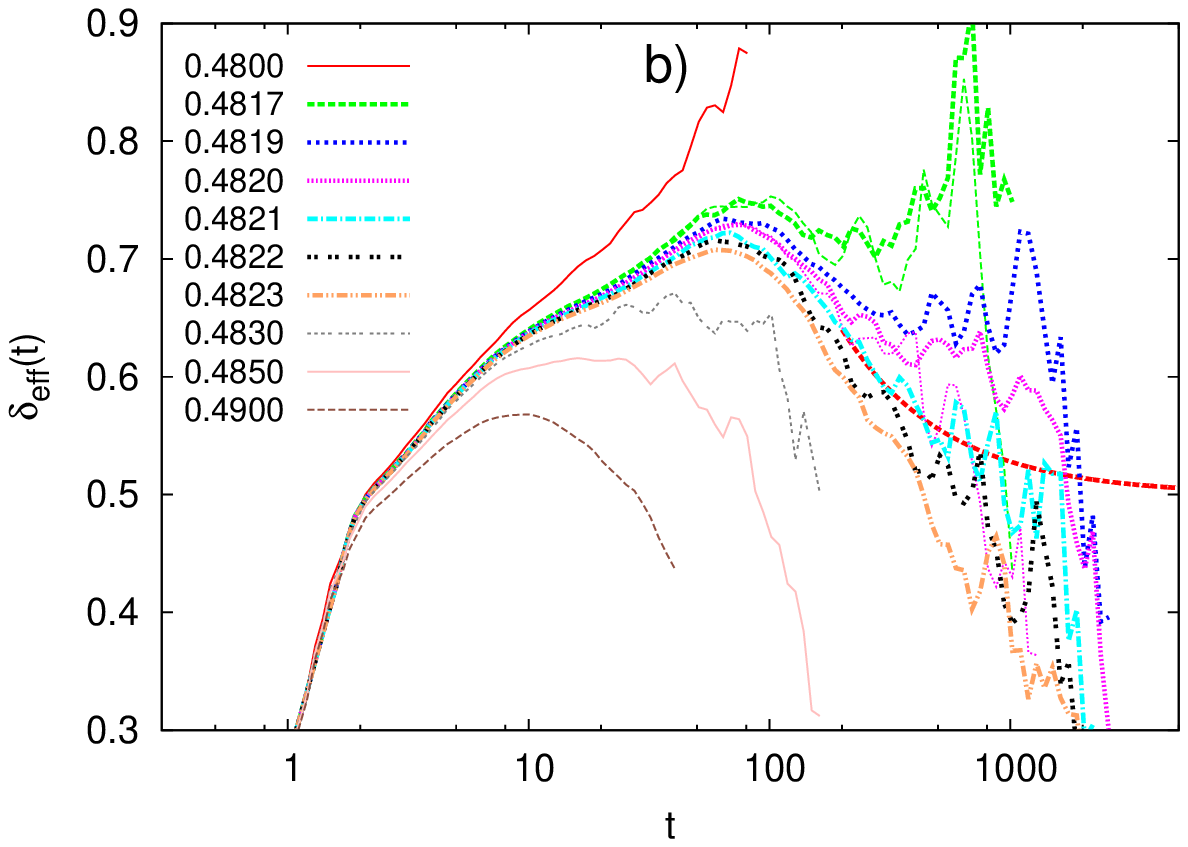}
\includegraphics[width=10cm]{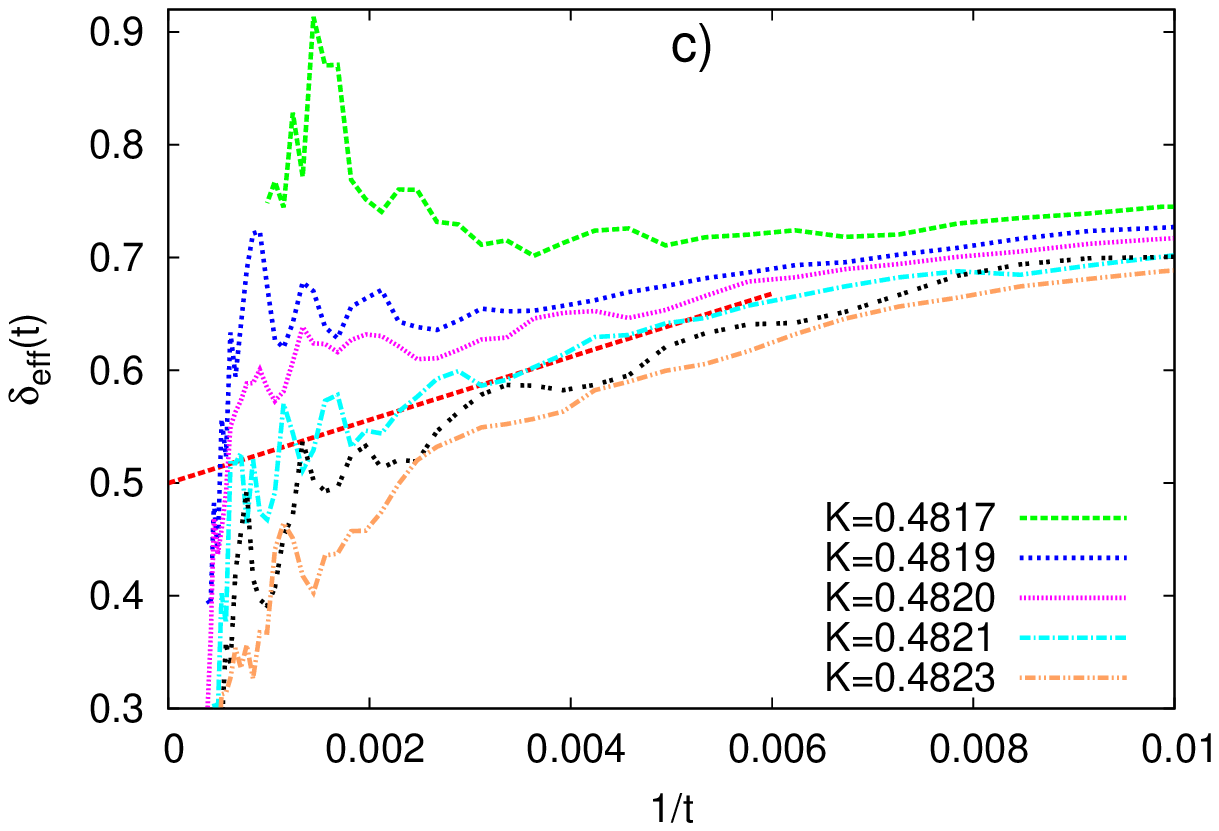} 
%erkm4+.ps
%erkm4tp.ps
%erkm4tp+.ps
\caption{\label{fig_delta} a) Time-dependence of the average order parameter obtained numerically for the Kuramoto model on ER graphs for different values of the coupling strength $K$.
b) Effective decay exponents $\delta_{\rm eff}(t)$ as a function of time, for different values of $K$. For data in the range $K=0.4817-0.4823$ the size was $N=2^{27}$ (thick lines) and  $N=2^{25}$ (thin lines), whereas for other values of $K$, $N=2^{22}$. The dashed red line has the form $\delta_{\rm eff}(t)=28/t+0.5$.
c) Effective decay exponents plotted against $1/t$.    
}
\end{center}
\end{figure}

The dependence of the average order parameter on time and that of the corresponding effective decay exponent, when the system is started from a fully synchronized state, are shown in Fig. \ref{fig_delta}. 
%%%%%%%%%%%%%%%%%%%%%%%%%%%%%%%%%%%%%%%%%%%%%%%%%%%%%%%%%%%%%%%%%%%%%%%%%
%%%%%%%%%%%%%%%%%%%%%%%%%%%%%%%%%%%%%%%%%%%%%%%%%%%%%%%%%%%%%%%%%%%%%%%%
As a comparison it is worth invoking the behavior of $\delta_{\rm eff}(t)$ in the Kuramoto model with all-to-all coupling, presented in Fig. 1 of Ref. \cite{choi}, in which model $K_c$ is exactly known. 
In that case, at the critical point, $\delta_{\rm eff}(t)$ increases monotonically with $t$ and saturates to $1/2$. For $K>K_c$, it increases initially, then, after reaching a maximum (with a maximum value below $1/2$), it starts to decrease to zero, while for $K<K_c$, it increases all the way up to its finite-size cutoff, having an inflection point around the saturation value of the critical curve, where they change from concave to convex.
  
As can be seen in Fig. \ref{fig_delta}b, $\delta_{\rm eff}(t)$ shows a qualitatively different behavior on ER graphs. 
For values of $K$ close to the transition, $\delta_{\rm eff}(t)$ is an increasing function for short times until $t\sim 10^2$, where it reaches a maximal value above $0.7$ and decreases afterwards, so that the asymptotic value of $\delta$ is approached from above. Note that the rapid drop of the near-critical curves for long times is due to the finite-size of the system. 
The curve $\delta_{\rm eff}(t)$ for $K=0.4817$ shows a steep increase (up to the cutoff) meaning that this point is in the unsynchronized phase, i.e. 
$K_c>0.4817$. On the other hand, the curve for $K=0.483$ has a concave shape with a rapid decrease at long times, meaning that it belongs to the synchronized phase, i.e. $K_c<0.483$. Whithin this relatively narrow range, the curves $\delta_{\rm eff}(t)$ vary sensitively with $K$, allowing us only for a rather inaccurate estimate on the asymptotic value, $\delta$, leaving room for a possible tendency toward the mean-field value, $1/2$, see the guide-to-the-eye at $K=0.4821$ in Fig.  \ref{fig_delta}b, or a tendency toward some higher limiting value for smaller values of $K$.   

\begin{figure}[ht]
\begin{center}
\includegraphics[width=10cm]{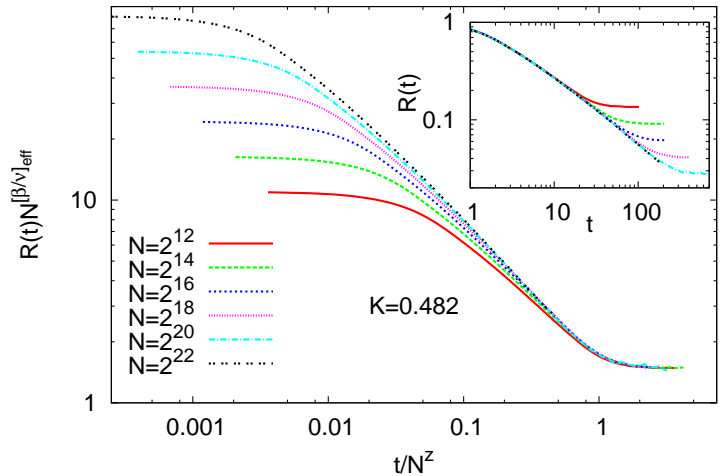} 
%erkm4.ps
\caption{\label{fig_dyn_scale} Scaling plot of the time-dependence of the average order parameter obtained numerically for the Kuramoto model on ER graphs using a fully synchronized initial state for different sizes at $K=0.482$. The parameters $[\beta/\overline{\nu}]_{\rm eff}=0.29$ and $\overline{z}=0.4$ are used. The inset shows the unscaled data.}
\end{center}
\end{figure}

\begin{figure}[ht]
\begin{center}
\includegraphics[width=10cm]{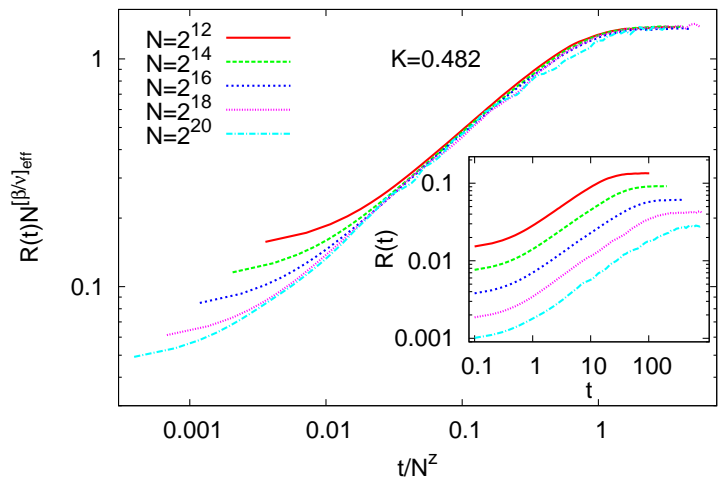} 
%erkm4ri.ps
\caption{\label{fig_ri} The same as in Fig. \ref{fig_dyn_scale} but with random initial states.}
\end{center}
\end{figure}

Having an estimate of the critical coupling strength, one usually determines the critical exponents by using the scaling relation in Eq. (\ref{scaling}). 
Setting $\Delta=0$ and $b=N$, we obtain that, having the time-dependent order parameter for different sizes and plotting $R(t,N)N^{\beta/\overline{\nu}}$ vs. $tN^{-\overline{z}}$ a data collapse is achieved provided the correct exponents are used. Regarding the dynamical decay exponent $\delta_{\rm eff}(t)$ shown in Fig. \ref{fig_delta}, the numerically available time scales are far from the asymptotic regime.
Nevertheless, the effective exponent $\delta_{\rm eff}(t)$ varies slowly in time, and in a limited range of time, which corresponds to a size range through $N\sim t^{1/\overline{z}}$, it can be regarded as roughly constant. Here, an approximate scaling collapse is still expected to hold provided effective values of the critical exponents are used. 
This is illustrated in Figs. \ref{fig_dyn_scale} and \ref{fig_ri}, for synchronized and random initial conditions, respectively, where the time scale corresponding to the applied sizes is near the time where $\delta_{\rm eff}(t)$ is maximal.
Here, satisfying data collapses are obtained with the values $\overline{z}=0.4$, which agrees with the mean-field value and $[\beta/\overline{\nu}]_{\rm eff}=0.29$, which is higher than the mean-field value $1/5$. 
%%%%%%%%%%%%%%%%%%%%%%%%%%%%%%%%%%%%%%%%%%%%%%%%%%%%%%%%%%%%%%%%%%%%%%%
%%%%%%%%%%%%%%%%%%%%%%%%%%%%%%%%%%%%%%%%%%%%%%%%%%%%%%%%%%%%%%%%%%%%%%%%

%%%%%%%%%%%%%%%%%%%%%%%%%%%%%%%%%%%%%%%%%%%%%%%%%%%%%%%%%%%%%%%%%%%%%%%%%
%%%%%%%%%%%%%%%%%%%%%%%%%%%%%%%%%%%%%%%%%%%%%%%%%%%%%%%%%%%%%%%%%%%%%%%%

\begin{figure}[ht]
\begin{center}
\includegraphics[width=10cm]{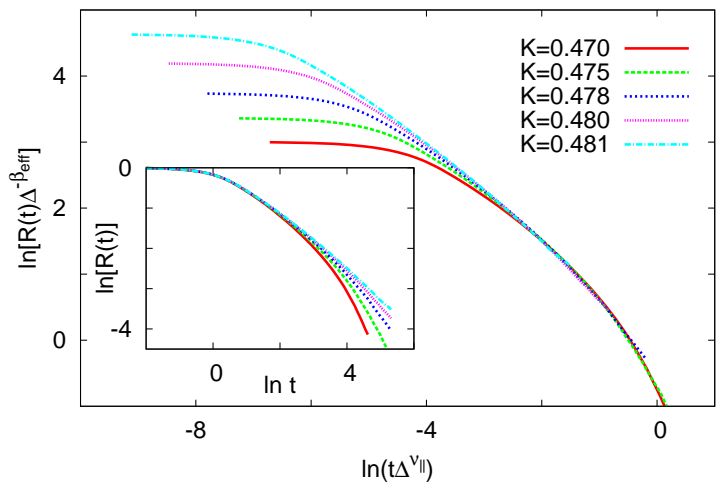} 
%erkm4off+.ps
\caption{\label{off}
Scaling plot of the time-dependent order parameter of the Kuramoto model on ER graphs obtained numerically with a fully synchronized initial state, for different coupling strengths in the unsynchronized phase. The system size was $N=2^{22}$ and the parameters $K_c=0.4821$, $\beta=0.68$, and $\nu_{\parallel}=1$ were used. The inset shows the unscaled data. 
}
\end{center}
\end{figure}

%%%%%%%%%%%%%%%%%%%%%%%%%%%%
%%%off-critical%%%%%%%%%%%%%
%%%%%%%%%%%%%%%%%%%%%%%%%%%%
The above methods provided estimates on the ratios of the (effective) static exponents $\beta$, $\overline{\nu}$, and $\nu_{\parallel}$. 
In order to determine their absolute values, we calculated the time-dependent order parameter for different coupling strengths in the unsynchronized phase. According to Eq. (\ref{scaling}), for system sizes well beyond the correlation size, plotting 
$R(t,\Delta)\Delta^{-\beta}$ against $t\Delta^{\nu_{\parallel}}$ results in a data collapse for long times. 
As can be seen in Fig. \ref{off}, this is indeed the case, and an approximate collapse is achieved by an effective order-parameter exponent $\beta_{\rm eff}=0.68(3)$ and a correlation-length exponent $\nu_{\parallel}=1.00(2)$ compatible with the mean-field value.    
%%%%%%%%%%%%%%%%%%%%%%%%%%%%%%%%%%%%%%%%%%%%%%%%%%%%%%%%%%%%%%%%%%%%%%%%%
%%%%%%%%%%%%%%%%%%%%%%%%%%%%%%%%%%%%%%%%%%%%%%%%%%%%%%%%%%%%%%%%%%%%%%%%

%%%%%%%%%%%%%%%%%%%%%%%%%%%%%%%%%%%%%%%%%%%%%
%%%%finite-size scaling
%%%%%%%%%%%%%%%%%%%%%%%%%%%%%%%%%%%%%%%%%%%%%%
To estimate the exponents, we also applied the finite-size scaling of the order parameter in the stationary state. In order to do this, we produced stationary data for different sizes and, for each $N$, for a series of different values of $K$, with a resolution $\Delta K=0.01$. 
To reduce the relaxation time, for each sample, the end state of the simulation at a given value of $K$ was used as the initial state for $K+\Delta K$. 
For the largest system size $N^{17}$, the simulation time was $2^{15}\cdot 0.1$ for each $K$, and measurement of the order parameter was performed in the last quarter of the simulation period.    

\begin{figure}[h]
\begin{center}
\includegraphics[width=10cm]{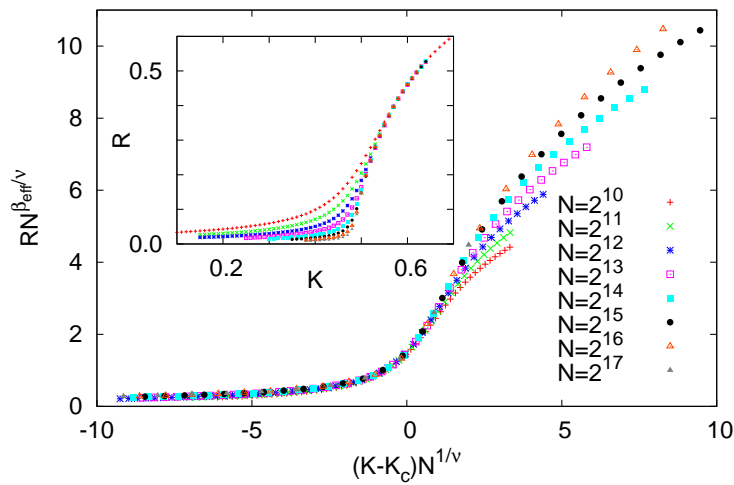} 
%oper.ps
\caption{\label{oper}
Scaling plot of the stationary order parameter of the Kuramoto model on ER graphs obtained numerically for different sizes and coupling strengths. The parameters $K_c=0.4821$, $\beta_{\rm eff}=0.72$, and $\overline{\nu}=2.5$ were used. The inset shows the unscaled data. 
}
\end{center}
\end{figure}

A scaling plot of the stationary order parameter according to Eq. (\ref{scaling}) with $b=N$ and $t\to\infty$ around the critical point is shown in Fig.~\ref{oper}. We find that the optimal collapse can be achieved with an effective order-parameter exponent higher than the mean-field value, $\beta_{\rm eff}=0.72$, and $\overline{\nu}=2.5$ which is compatible with the mean-field universality class. %%%%%%%%%%%%%%%%%%%%%%%%%%%%%%%%%%%%%%%%%%%%%%%%%%%%%%%%%%%%%%%%%%%%%%%%%
%%%%%%%%%%%%%%%%%%%%%%%%%%%%%%%%%%%%%%%%%%%%%%%%%%%%%%%%%%%%%%%%%%%%%%%%

\begin{figure}[ht]
\begin{center}
\includegraphics[width=10cm]{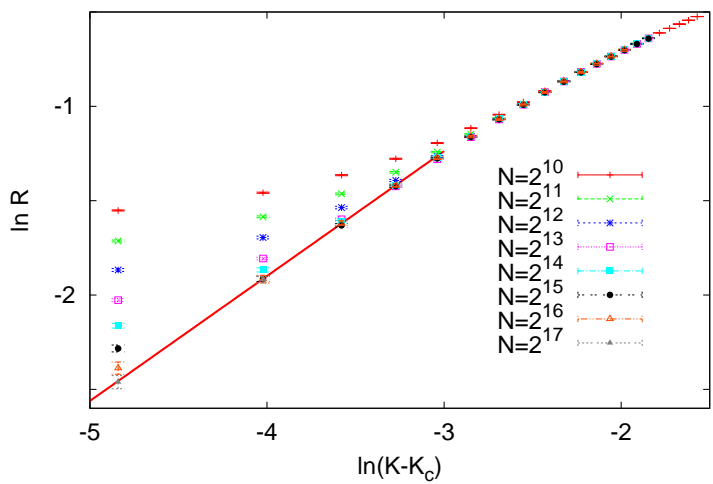}
%beta.ps 
\caption{\label{beta}
The logarithm of the average stationary order parameter plotted against the logarithmic distance from the estimated critical point $K_c=0.4821$ for different system sizes. The straight line has the slope $0.66$.
}
\end{center}
\end{figure}

According to Eq. (\ref{scaling}), the stationary order parameter vanishes in the synchronized phase close to the critical point as 
\be
R(\Delta)\sim \Delta^{\beta},
\ee
which is valid for an infinite system. This is tested in Fig.~\ref{beta}, where the estimate of the critical coupling strength $K_c=0.4821$ has been used.
%%%%%%%%%%%%%%%%%%%%%%%%%%%%%%%%%%%%%%%%%%%%%%%%%%%%%%%%%%%%%%%%%%%%%%%%%
%%%%%%%%%%%%%%%%%%%%%%%%%%%%%%%%%%%%%%%%%%%%%%%%%%%%%%%%%%%%%%%%%%%%%%%%
As can be seen, the effective order-parameter exponent 
$\beta_{\rm eff}(\Delta)=d\ln[R(\Delta)]/d\ln\Delta$ slowly changes with $\Delta$. For the data points close to the critical point, where a saturation of the dependence on $N$ is reached we have $\beta_{\rm eff}=0.66$, which is again well above the corresponding mean-field value $1/2$.

\begin{figure}[ht]
\begin{center}
\includegraphics[width=10cm]{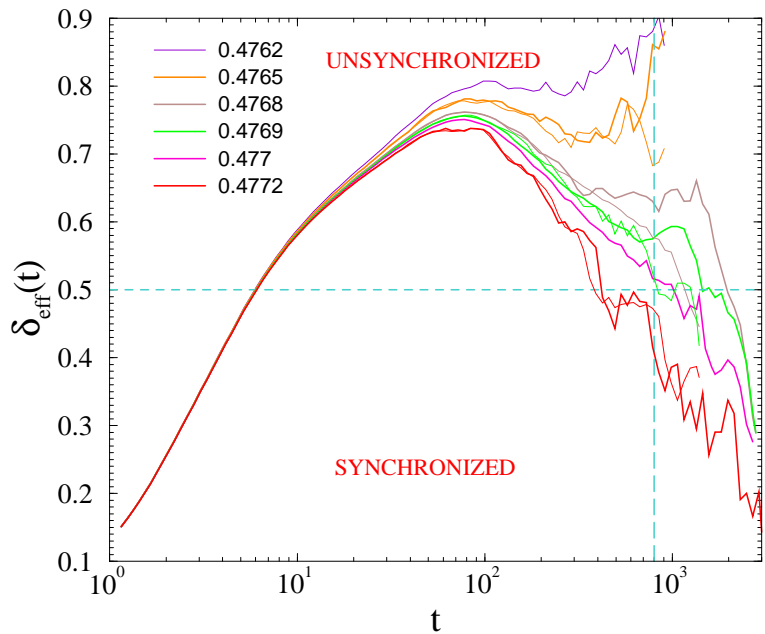}
%sl_ll-3.eps
\caption{\label{fig_er2d}
Effective decay exponents $\delta_{\rm eff}(t)$ as a function of time obtained numerically in the ER graph ($k=1$) combined with a 2d lattice, for different values of $K$ and sizes $N=6000$ (thin lines) and $N=10000$ (thick lines). The dashed horizontal line indicates the asymptotic value $\delta=1/2$ of the mean-field universality class, while the vertical line shows the estimated cutoff time for the larger size.  
}
\end{center}
\end{figure}

Besides the ER graph we also studied the Kuramoto model on a union of ER graphs with a regular two dimensional lattice. The results are qualitatively similar to those obtained on ER graphs. The effective decay exponents $\delta_{\rm eff}(t)$ obtained for combinations of $k=1$ graphs with a 2d lattice are shown in 
Fig.~\ref{fig_er2d}. 
%%%%%%%%%%%%%%%%%%%%%%%%%%%%%%%%%%%%%%%%%%%%%%%%%%%%%%%%%%%%%%%%%%%%%%%%%
%%%%%%%%%%%%%%%%%%%%%%%%%%%%%%%%%%%%%%%%%%%%%%%%%%%%%%%%%%%%%%%%%%%%%%%%
The critical coupling strength is estimated to be in the range 
$0.4765<K_c<0.4772$. But within this range, the effective exponents change significantly even up to the cutoff time, allowing for a possible approach to the mean-field value $1/2$ at later times, as well as a saturation to a limit higher than $1/2$.

%%%%%%%%%%%%%%%%%%%%%%%%%%%%%%%%%%%%%%%%%%%%%%%%%%%%%%%%%%%%%%%%%%%%%%%%%%%
\section{Discussion and outlook}
\label{sec:discussion} 

We have seen in the previous section that the estimates of the correlation-size exponent $\overline{\nu}$, the temporal correlation-length exponent $\nu_{\parallel}$ and their ratio $\overline{z}$ are compatible with the mean-field universality class. The order-parameter exponent $\beta$ as well as the critical exponents involving it are, however, loaded with strong corrections, 
which, inspite of the extremely large sizes attained by an enormous GPU resource usage, do not allow for a clear conclusion on the asymptotic value. 

A more general form of critical scaling contains a power-law correction to the leading term \cite{essam}, such as  
\be 
R(t)\sim t^{-\delta}(1+bt^{-\delta'}) 
\ee
in case of the time-dependence of the order parameter.  
This provides an algebraic correction to the effective exponent 
$\delta_{\rm eff}(t)=-d\ln R(t)/d\ln t$ as: 
\be 
\delta_{\rm eff}(t)=\delta+b\delta't^{-\delta'}.
\label{delta_corr}
\ee
A first guess for the correction exponent is $\delta'=1$, which we found numerically also in the Kuramoto model with all-to-all coupling (not shown), and it is worth mentioning that this form of the correction has been observed also in the contact process \cite{essam}. 
A test of this type of corrections can be seen in Fig.  \ref{fig_delta}c, where  $\delta_{\rm eff}(t)$ is plotted against $1/t$ and the near-critical curves are compatible with a linear dependence.  

In accordance with the numerical results presented in Fig. \ref{fig_dyn_scale}, due to the correction in $\delta$, corrections also appear in the exponent $\beta/\overline{\nu}$ describing the finite-size scaling of the stationary order parameter in the transition point through $R(N)\sim N^{-\beta/\overline{\nu}}$. Its effective value is related to that of exponent $\delta$ according to 
\be 
[\beta/\overline{\nu}]_{\rm eff}(N)\approx \delta_{\rm eff}[\tau(N)]\overline{z},
\ee
where $\tau(N)\sim N^{\overline{z}}$ is the relaxation time needed to reach the stationary state in finite systems of size $N$. 
Using Eq. (\ref{delta_corr}), this leads to  
\be 
[\beta/\overline{\nu}]_{\rm eff}(N)\approx 
\beta/\overline{\nu} + const\cdot N^{-\delta'\overline{z}}. 
\ee

\begin{figure}[ht]
\begin{center}
\includegraphics[width=10cm]{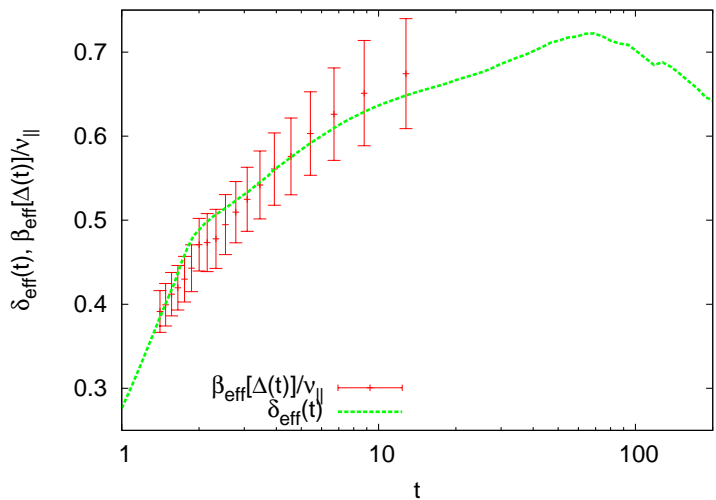} 
%betaeff.ps
\caption{\label{betaeff} The effective decay exponent $\delta_{\rm eff}(t)$ at $K=0.4821$ compared to the effective order-parameter exponent $\beta_{\rm eff}[\Delta(t)]/\nu_{\parallel}$ as a function of the equivalent time corresponding to the control parameter $\Delta$ through $t=A\Delta^{-\nu_{\parallel}}$ with $A=3.5$. 
}
\end{center}
\end{figure}

The effective order-parameter exponent, $\beta_{\rm eff}(\Delta)=d\ln[R(\Delta)]/d\ln\Delta$, is related to the effective decay exponent through 
\be 
\beta_{\rm eff}(\Delta)\approx \delta_{\rm eff}[\tau(\Delta)]\nu_{\parallel},
\label{deltabeta} 
\ee
where 
\be 
\tau\simeq A\Delta^{-\nu_{\parallel}}
\label{taudelta}
\ee
is the correlation time at a distance $\Delta$ from the transition point. 
Using Eq. (\ref{delta_corr}), this yields 
\be 
\beta_{\rm eff}(\Delta)\approx \beta + const\cdot \Delta^{\delta'\nu_{\parallel}}
\label{eqbetaeff}
\ee
close to the transition point. 
We have calculated $\beta_{\rm eff}(\Delta)$ from the discrete data points presented in Fig. \ref{beta} for which size-dependence is negligible, and compared it to $\delta_{\rm eff}(t)$ by using an equivalent time corresponding to the control parameter through Eq. (\ref{taudelta}). 
As shown in Fig. \ref{betaeff}, the approximate relationship in Eq. (\ref{deltabeta}) is satisfactorily fulfilled with an appropriate choice of the constant $A$.   
%%%%%%%%%%%%%%%%%%%%%%%%%%%%%%%%%%%%%%%%%%%%%%%%%%%%%%%%%%%%%%%%%%%%%%%%%
%%%%%%%%%%%%%%%%%%%%%%%%%%%%%%%%%%%%%%%%%%%%%%%%%%%%%%%%%%%%%%%%%%%%%%%%
We can also see that $\beta_{\rm eff}(\Delta)$ for the studied range of $\Delta$ has not even reached its maximum and it is therefore well out of the range of
validity of Eq. (\ref{eqbetaeff}).

We have demonstrated in this work that, in the Kuramoto model on sparse random graphs, the scaling of the order parameter is loaded with strong corrections, due to which the estimates of critical exponents involving the order-parameter exponent obtained by numerical studies of finite systems may significantly differ from their asymptotic values. 
A usual way of checking the compatibility of the Kuramoto model on graphs with the mean-field universality class is to judge whether a good scaling collapse of finite-size stationary order parameter data can be achieved by using mean-field critical exponents and an appropriate choice of the critical coupling strength $K_c$. This has been done in the case of the ER graph with $k=4$, the same model studied here, in Ref. \cite{um}, where a scaling collapse was obtained up to the size $N\approx 2^{16}$ by using $K_c=0.49$. 
As can be read off from Fig. \ref{fig_delta}b, the effective exponents $\delta_{\rm eff}(t)$ for $K=0.49$ are, for times corresponding to the applied sizes, indeed close to the mean-field value $0.5$. 
Nevertheless forcing the asymptotic value of the critical exponents to finite-size data, for which the corrections are still significant, leads to a considerable shift in the estimation of $K_c$. 

We have seen in this work that the finite-size corrections for the critical scaling of the Kuramoto model are stronger for sparse random graphs than for the variant with all-to-all coupling. This may be generally valid for other models as it is related to the sparse interaction structure, but it is particularly important for the Kuramoto model as the size scales available by numerical integration of the dynamical equations are well below those available by simulations of stochastic reaction-diffusion systems.  
We have seen that also the form of corrections is qualitatively different for sparse graphs and the case of all-to-all coupling, the effective exponents tending to the asymptotic value from above and from below, respectively. 
An interesting question is how the effective exponents behave for random graphs having less edges than the all-to-all coupling case [$N(N-1)/2$] but more than sparse graphs with a constant mean degree [$O(N)$] have. 
Similarly, one can ask whether such an anomalous, non-monotonic behavior of the effective exponents appears for the Kuramoto model on finite dimensional lattices. These problems are left for future research.

%%%%%%%%%%%%%%%%%%%%%%%%%%%%%%%%%%%%%%%%%%%%%%%%%%%%%%%%%%%%%%%%%%%%%%%%%%%%%
%%%%%%%%%%%%%%%%%%%%%%%%%%%%%%%%%%%%%%%%%%%%%%%%%%%%%%%%%%%%%%%%%%%%%%%%%%%%%

\ack
This work was supported by the Hungarian Scientific Research Fund under grant
No. K128989.
We gratefully acknowledge computational resources provided by NIIF Hungary, the
HZDR computing center, the Group of M. Bussmann and the Center for Information
Services and High Performance Computing (ZIH) at TU Dresden via the GPU Center
of Excellence Dresden. We thank S. Gemming for support.


\section*{References}
\begin{thebibliography}{99}

\bibitem{pikovsky}
Pikovsky A, Rosenblum M, Kurths J 2001 {\it Synchronization: A Universal Concept in Nonlinear Science} (Cambridge University Press, Cambridge) 

\bibitem{acerbon} Acebr\'on J A, Bonilla L L, P\'erez Vicente C J, Ritort F, Spigler R 2005 {\it Rev. Mod. Phys.} {\bf 77} 137

\bibitem{arenas} Arenas A, Diaz-Guilera A, Kurths J, Moreno Y, and Zhou C S
2008 {\it Phys. Rep.} {\bf 469} 93

\bibitem{kuramoto} Kuramoto Y 1984 in {\it Proceedings of the International Symposium on Mathematical Problems in Theoretical Physics} edited by Araki H (Springer, New York);  1984 {\it Chemical Oscillations, Waves, and Turbulence}
(Springer, Berlin)

\bibitem{oa}
Ott E, Antonsen T M 2008 {\it CHAOS} {\bf 18} 037113

\bibitem{chate_prl}
Hong H, Chat\'e H, Park H, and Tang L-H 2007 
{\it Phys. Rev. Lett.} {\bf 99} 184101


%%%%%%regular%%%%%%%
\bibitem{sakaguchi}
Sakaguchi H, Shinomoto S, and Kuramoto Y 1987
{\it Prog. Theor. Phys.} {\bf 77} 1005

\bibitem{hong2005} 
Hong H, Park H, and Choi M Y 2005
{\it Phys. Rev.} E {\bf 72} 036217 

%%%%%LR interaction%%%%%%%
\bibitem{rogers}
Rogers J L, Wille L T 1996
{\it Phys. Rev.} E {\bf 54} R2193

\bibitem{vicsek}
Mar\'odi M, d'Ovidio F, and Vicsek T 2002 
{\it Phys. Rev.} E {\bf 66} 011109

\bibitem{chowdhury}
Chowdhury D, Cross M C 2010
{\it Phys. Rev.} E {\bf 82} 016205

\bibitem{uchida}
Uchida N 2011
{\it Phys. Rev. Lett.} {\bf 106} 064101

%%%%%SW network%%%%%
\bibitem{hong2002}
Hong H, Choi M Y, and Kim B J 2002 
{\it Phys. Rev.} E {\bf 65} 026139 

%%%%%%%ER-graph%%%%
\bibitem{um}
Um J, Hong H, and Park H 2014 
{\it Phys. Rev.} E {\bf 89} 012810

%%%%%%%%%%%%%
\bibitem{bianconi}
Mill\'an A P, Torres J J, and Bianconi G 2019
{\it Phys. Rev.} E {\bf 99} 022307
%%%%%scaling%%%%%%%
\bibitem{lee}
Lee M J, Yi S D, and Kim B J 2014
{\it Phys. Rev. Lett.} {\bf 112} 074102

\bibitem{MunPNAS}
di Santo S, Villegas P, Burioni R, and Mu\~noz M A 2008
{\it Proc. Natl. Acad. Sci. U.S.A.} {\bf 115} E1356

\bibitem{choi}
Choi C, Ha M, and Kahng B 2013
{\it Phys. Rev.} E {\bf 88} 032126 


\bibitem{HMN} \'Odor G, Dickman R and \'Odor G 2015
{\it Sci. Rep.} 14451


%%%%%%%GPU%%%%%%%%%%%

\bibitem{KOGkuramotoGPU_tbp}
Kelling J, \'Odor G, and Gemming S \textit{to be published}


%%%%%%%CP correction%%%%%%%%%%%
\bibitem{essam}
Essam J W, De'Bell K, Adler J, and Bhatti F M 1986 
{\it Phys. Rev.} B {\bf 33} 1982



\end{thebibliography}
\end{document}